\documentclass{aastex6}

\usepackage{amsmath}
\usepackage{graphicx}
\usepackage{natbib}
\usepackage{mathrsfs}
\usepackage{multirow}
\usepackage{longtable}
\usepackage{url}

\begin{document}
\title{ FORMING DIFFERENT PLANETARY ARCHITECTURES . I . FORMATION EFFICIENCY OF HOT JUPITES FROM HIGH-ECCENTRICITY MECHANISMS}

\shorttitle{Formation of hot Jupiters by high-eccentricity migration}
\shortauthors{
WY,ZHOU JL}

\author{Ying Wang\altaffilmark{1}, Ji-lin Zhou\altaffilmark{1}, Liu hui-gen\altaffilmark{1} and Zeyang Meng\altaffilmark{1,2}
}

\altaffiltext{1}{
School of Astronomy and Space Science and Key Laboratory of Modern Astronomy and Astrophysics in Ministry of Education, Nanjing University, Nanjing 210093, China; zhoujl@nju.edu.cn
}
\altaffiltext{2}{Department of Astronomy and Astrophysics, University of California, Santa Cruz, CA 95060, USA}
\email{zhoujl@nju.edu.cn}

\begin{abstract}
Exoplanets discovered over the last decades have provided a new sample of giant exoplanets, hot Jupiters. For lack of enough materials in current locations of hot Jupiters, they are perceived to form outside snowline. Then, migrate to the locations observed through interactions with gas disks or high-eccentricity mechanisms. We examined the efficiencies of different high-eccentricity mechanisms to form hot Jupiters in near coplaner multi-planet systems. These mechanisms include planet-planet scattering, Kozai-Lidov mechanism, coplanar high-eccentricity migration, secular chaos, as well as other two new mechanisms we find in this work, which can produce hot Jupiters with high inclinations even retrograde. We find Kozai-Lidov mechanism plays the most important role in producing hot Jupiters among these mechanisms.  Secular chaos is not the usual channel for the formation of hot Jupiters due to the lack of angular momentum deficit within ${10^7}$$T_{\rm in}$ (periods of the inner orbit). According to comparisons between the observations and simulations, we speculate that there are at least two populations of hot Jupiters. One population migrates into the boundary of tidal effects due to interactions with gas disk, such as ups And b, WASP-47 b and HIP 14810 b. These systems usually have at least two planets with lower eccentricities, and keep dynamical stable in compact orbital configurations. 
Another population forms through high-eccentricity mechanisms after the excitation of eccentricity due to dynamical instability. 
This kind of hot Jupiters usually has Jupiter-like companions in distant orbits with moderate or high eccentricities.

\end{abstract}


\vspace{1cm}

\section{INTRODUCTION}
\label{sc:intro}
Since the first exoplanet (51 Pg b) around solar-like stars was discovered in 1995 \citep{Mayor1995}, the number of confirmed exoplanets has increased to 3483, in 2600 exoplanet systems. Among them, there are 581 multi-planet systems \footnote{http://exoplanetarchive.ipac.caltech.edu/index.html}.
Due to limits of detections, the observed multi-planet systems may not be complete. Yet they have shown various orbital architectures.  51 Pg b is the first hot Jupiter observed. Then astronomers found orbital orientations of some hot Jupiters and spin axes of their hosts are not aligned \citep{Fabrycky2009}, even retrograde hot Jupiters \citep{Winn2009}. 
Planetary formation and evolution can last as long as 10~Gyrs with various procedures.  According to the classical core accretion theory,  planet embryos form through ``runaway" and ``oligarchic growth" from planetesimals in a proto-stellar disk, enter into the process of quickly gas accretion and then become gas giants if there are enough gas \citep{Ida2004,Armitage2007}.
Simultaneous or subsequent processes such as disk evolution and gas dissipation with time \citep{Lynden-Bell1974,Hayashi1981,Larwood1996},  planet migration in gas disk \citep{Goldreich1980, Lin1986, Lin1996, Ward1997,Cossou2013}, and dynamical instability of multi-planet systems \citep{Rasio1996,Chambers1996,Ford2001,Marzari2002,Chatterjee2008,Ford2008,Nagasawa2008,Smith2009,Funk2010,Guillochon2011,Beauge2012} might occur,  which could lead to complicated planetary configurations as observations.  As we all know, multi-planet systems are non-integrable and chaotic. Thus, to understand the typical orbital configurations of exoplanets, statistical studies are needed.
  
In this series of papers, we study the formation of planetary architectures in multi-planet systems in order to answer the question: what governs the formation and evolution of the planetary architecture. In the present paper, we revisit the formation of Hot Jupiters (HJs) through high-eccentricity mechanisms. Planets with mass $13{M_J}\ge m ( \hspace{0.15cm} or \hspace{0.15cm} m\sin i) \ge 0.3{M_J}$, orbital period $P \le 10$ days are recognized as HJs in this article. We collected data of HJs from websites: \url{http://exoplanet.eu/}, \url{http://exoplanets.org/}, \url{http://exoplanetarchive.ipac.caltech.edu/}, \url{http://openexoplanetcatalogue.com/}, \url{http://www.astro.keele.ac.uk/jkt/tepcat/rossiter.html}. To date, the number of HJs has increased to 351. The catalog of corresponding systems are listed in website: \\
\url{https://github.com/astro-WangYing/Hot-Jupiters-catalog}. The main characteristics of HJs are: 
\begin{enumerate}
\item HJs are mostly single ones, at least without massive close neighbours.  Only 11 hot Jupiters have been detected having planetary companions, their semi-major axes range from 0.5 au to 5.6 au. The information of these systems are listed in Table \ref{tab:HJC}.
\item 332 HJs were detected in single star planetary systems, while 19 HJs in multi-star systems with S-type orbits.
\item 10 HJs in single star planetary systems and 2 in multi-star systems have retrograde orbits.
\end{enumerate}

Through Rossiter-McLaughlin effect, the misalignments between the orbital orientations of some HJs and spin axis of their host stars were discovered. The formation mechanisms of HJs' high obliquity can be summerized into two kinds. 
The first kind is ``stellar misalignments". The planetary disk spin axis is strongly misaligned with the stellar rotation axis, for the formation of star is an inherently chaotic process with variable accretion in the collapse of a turbulent molecular cloud. The final disc rotation axis depends on what fell into central star in the last \citep{Bate2010,Thies2011}. Spin orientation of magnetic host star can be pushed away from the disc spin axis towards the perpendicular state and even retrograde state through interaction between the star and circumstellar disc \citep{Lai2011, Foucart2011}, or stellar surface modulation on slow spin stars by internal gravity waves \citep{Rogers2012, Rogers2013a, Rogers2013b}.
The second kind is ``planetary misalignments". HJs can obtain high inclinations during the processes of dynamical interactions. This includes violent scattering between planets \citep{Rasio1996, Nagasawa2008, Nagasawa2011, Beauge2012} ; Kozai-Lidov mechanism \citep{Kozai1962, Lidov1962} induced by planets \citep{Naoz2011}, stellar companion\citep{Wu2003, Naoz2012, Dong2014, Petrovich2015b} or gas disk \citep{Terquem2010}; Kozai-Lidov mechanism after chained secular resonance \citep{Chen2013}; secular chaos in the system with enough angular momentum deficit \citep{Wu2011}.  The high eccentricities obtained from these mechanisms can be damped by stellar tides when the inner planets close enough to the hosts. Then elliptical orbits will be circularized, and these planets become HJs.
Except planet-planet scattering, all the scenarios of ``dynamical misalignments" to form retrograde HJs have some requirements on initial conditions \citep{Wu2011, Naoz2011}. So planet-planet scattering is the most possible route to form HJs in multi-planet systems without invoking the presence of binary companion \citep{Wu2003} or massive disk \citep{Terquem2010,Chen2013}. The compact spaces required to trigger dynamical instability of nearby gas giants are easily reach during the process of rapidly gas accretion in classical core accretion theory, or during the process of disk migration or gas dissipation.

Dynamical instabilities between two giants came up to explain the formation of first series of exoplanets \citep{Rasio1996}. However, owing to lack of fierce interactions between two giants, final inclinations of inner orbits are smaller than ${15^ \circ }$ \citep{Ford2001} which can't explain high obliquities of observed HJs. \citet{Marzari2002} studied the statistical outcomes  of remained inner planets after dynamical instability in three-giant systems with fixed initial semi-major axes, and found the inclinations of inner planets distributed in a large range, even retrograde. The paper predicted HJ formed through ``Jumping Jupiters" model should have another planet of comparable mass in a distant orbit.  Adopting the same initial conditions with \citet{Marzari2002}, \citet{Nagasawa2008} stated HJs can form due to short Kozai-Lidov state and tidal circularization during the procedure of dynamical instability before ejection of any planet. And the possibility of these HJs is twice the ones formed in long term stable after ejection. However, \citet{Beauge2012} simulated multi-planet systems consist of three or four giants located at unstable mean motion resonances, including tidal and relativistic effects as well as precession due to stellar oblateness. This research showed the population of HJs which formed in the systems where no planet ejected is transient. For most of these HJs hit with the central star within 1 Gyr. And, the population of HJs formed in systems with at least one planet ejected out can survive longer than 1 Gyr. \citet{Naoz2012} studied the evolution of Jupiter-like planets in stellar binaries through Monte Carlo simulations, and pointed out `` eccentric Kozai-Lidov mechanism" play an important role in the HJs formation in stellar binaries. To sum up, previous studies mainly focused on the total efficiencies of HJs formation during dynamical evolutions of multi-planet systems and fixed space separations. We want to find out respective roles of planet-planet scattering and other high-eccentricity mechanisms playing in the formation of HJs, and which is the dominant one. In addition, we also want to know the influences of the planet number, space separation and location of inner planet on the final HJs system. 

In this article, we check the efficiencies of HJs formation after dynamical instability of near coplaner multi-planet systems with various initial conditions. The article includes 5 sections. Numerical models and initial conditions, as well as methods to distinguish different dynamical mechanisms are shown in Section \ref{sec:szmn}. The statistical results about efficiencies of HJs formation through different dynamical mechanisms, the relations with planet number, space separation and location of inner planet are listed in Section \ref{sec:szjg}. Then, we compare retrograde proportion and $a \sim e$ scatter diagram of planetary companions of hot Jupiters in observations with our simulations in Section \ref{sec:observation}. The last Section shows conclusions and discussions.

%
%
\section{MODELS AND CRITERIA}\label{sec:szmn}
\subsection{Numerical Model and Initial Conditions}\label{sec:model}

In this article, we use `EMS' systems to simulate the evolutions of multi-planet systems, where $N$ planets have equal mass $m$ and equal scaled separations moving around a star with mass ${M_{*} }$ \citep{Zhou2007}. The space separation is defined by
\begin{equation}
k = \frac{{{a_{n + 1}} - {a_n}}}{{{R_{\rm H}}}}(n = 1,2,3, \ldots, N),
\label{eq:k}
\end{equation}
${a_n}$ is the semi-major axes of the $n$th planet, ${R_{\rm H}}$ is the mutual Hill radius,
\begin{equation}
{R_H} = {\left( {\frac{{2\mu }}{3}} \right)^{{1 \mathord{\left/
 {\vphantom {1 3}} \right.
 \kern-\nulldelimiterspace} 3}}}\frac{{{a_n} + {a_{n + 1}}}}{2}.
\label{eq:RH}
\end{equation}
$\mu$ is the reduced mass of each planet, $\mu = {m \mathord{\left/
 {\vphantom {m {{M_{*} }}}} \right.
 \kern-\nulldelimiterspace} {{M_{*} }}}$.

Our multi-planet systems consist of 2-5 planets with $\mu = {10^{ - 3}}$. The space separations between adjacent planets are in the range $2 \le k \le 6$ with interval $\Delta k = 0.001$.  Initial orbits of all planets are circular, and inclinations obey rayleigh distributions, with expectation ${1^ \circ }$.  Other orbital angles are chosen randomly from 0 to $2\pi $. We select two groups of innermost semi-major axes $a_{\rm in}$, 1 au and 5 au, and use Bulirsch-Stoer (BS) algorithm \citep{Press1992} in Mercury code \citep{Chambers1999} which can accurately calculate the revolutions of high-eccentricity orbits and close encounters.  A planet which hit the central star or is ejected out of 1000 AU will be moved out of planetary system. Integrations will be terminated when integrate ${10^7}$$T_{\rm in}$ \citep{Nagasawa2008,Nagasawa2011}, or only one planet is remained with orbital energy  negative. $T_{\rm in}$ is the period of initial inner orbit. After integrate ${10^7}$$T_{\rm in}$, most of multi-planet systems in our simulations have finished fierce dynamical interaction and turned into secular interaction state. The simulations don't include general relativity and tidal effect. General relativity and tidal effects can influence the evolution of eccentricity of inner orbit, which may confuse or cover up the characteristics of dynamical mechanisms which the systems had gone through. In order to distinguish them, we just set a boundary of tidal effects rather than add tidal force into our simulations.  In the next section, we will give the criteria of each high-eccentricity  mechanism to form hot Jupiters.

\subsection{Criteria of Different Mechanisms to Produce HJs}\label{sec:condition}

There are at least four kinds of high-eccentricity mechanisms to form HJs in our simulations. They are planet-planet scattering, coplanar high-eccentricity migration, Kozai-Lidov mechanism, and secular chaos. Next, we will give the condition of each mechanism needed to produce a hot Jupiter.

1.  Planet-planet scattering (PPS).

In classical core accretion model of planetary formation, compact separations between giants triggering to instability could be easily achieved at the stage of gas accretion. Solid material feeding zone of each planet embryo or core is about $10{R'_H}$\citep{Ida2004}, where ${R'_H}$is Hill radius, ${{R'}_H} = {(\frac{\mu }{3})^{\frac{1}{3}}}a$. The typical mass of cores of Jupiter-like planets are about $10{M_ \oplus }$ \citep{Ida2004}. After gas accretion, the separation changed into $3.11{{R'}_H}$ (or $2.23{R_H}$) for Jupiter mass, giving rise to dynamical instability naturally. 
With the dissipation of gas and subsequently weakening of damp effect on the giants, dynamical instability can also take place. 

Yet, the planets undergone dynamical instability can further enter into many mechanisms. In this article, we classify the following situations as PPS: (i)  Only one planet 
survived has enough large eccentricity close to the host and can be circularized by tidal effects.  (ii) If more than one planet remained, the inner planet is nearly decoupled with the outer ones. And it's eccentricity large enough. (iii) During interactions among planets, the final survived planet can reach to the host closely enough for a certain time. Here, we roughly identify that tidal dissipation and circularization will operate when the inner planet close to the host star ${a_1}(1 - {e_1}) \le 0.05$ au \citep{Marzari2002,Wu2011}. 

2.  Coplanar high-eccentricity migration (CHEM).

\citet{Petrovich2015a} found, in a two-planet system, if mutual inclination $\lesssim {20^ \circ }$, and satisfy 
\begin{equation}
{e_1}\lesssim 0.1,{e_2} \gtrsim 0.67, \alpha \lesssim 0.3
\label{eq:CHEM1}
\end{equation}
or
\begin{equation}
{e_1},{e_2} \gtrsim 0.5, \alpha \lesssim 0.16,
\label{eq:CHEM2}
\end{equation}
here, $\alpha ={m_{1}}/{m_{2}}{({a_{1}}/{a_{2}})^{{1 \mathord{\left/
 {\vphantom {1 2}} \right.
 \kern-\nulldelimiterspace} 2}}}$, pericenter of inner planets can periodically close to the host, become HJs with inclination $\le {30^ \circ }$. 

In the end of simulations, if more than one planet is remained, and mutual inclination between inner and outer planet meets $ I_{mut} \le {20^ \circ }$, as well as satisfy Eq.(\ref{eq:CHEM1}) or (\ref{eq:CHEM2}), we regard the inner planet as a HJ. \citet{Li2014} show that coplanar high-eccentricity migration can make inner orbits flip when ${e_{1}}$ is extremely large. However, the author ignored the effects from tides and tidal disruptions, and regarded the planets as test particles which will greatly weaken the general relativity precession. Generally, CHEM is more likely to produces HJs with low inclination. Retrograde HJs might formed only if the semi-major axes of inner planet is very large ($\gg 1$ au) \citep{Petrovich2015a}. In our simulations, the inner planet is too close to the host to flip. So, in our paper, all of the HJs from CHEM are prograde. 

3.  Kozai-Lidov mechanism (Kozai).

In circular restricted hierarchical three-body system, if the mutual inclination between two planets is larger than $39^ \circ$, they will enter into Kozai-Lidov libration \citep{Kozai1962,Lidov1962}. If the eccentricity of inner planet can become large enough to make the pericenter into tidal effects of host in periodic variation, the combined action of Kozai-Lidov libration and tidal effects can produce a highly inclined hot Jupiter. If the mutual inclination between two planets is smaller than $39^ \circ$, they will enter into Kozai-Lidov circulation. And this can produce a near coplaner hot Jupiter. In our simulations, all of the planets have Jupiter mass, we should use general three-body criterion to distinguish Kozai-Lidov mechanism. {\bf In Jacobi coordinate system, adopting Delaunays elements, just considering the quadrupole approximation (when $\epsilon  = \left( {\frac{{M_{*} - {m_1}}}{{M_{*} + {m_1}}}} \right)\left( {\frac{{{a_1}}}{{{a_2}}}} \right)\left( {\frac{{{e_2}}}{{1 - e_2^2}}} \right) \le 0.001$\citep{Li2014chaos})}, analytical maximum eccentricity of inner planet (when ${\omega _1} = \frac{\pi }{2}$, ${\omega _1}$ is the augument of pericenter of inner planet ), can be obtained using conservation of total angular momentum and energy
\citep{Naoz2013},
\begin{equation}
L_1^2(1 - e_1^2) + 2{L_1}{L_2}\sqrt {1 - e_1^2} \sqrt {1 - e_2^2} \cos {i_{\rm tot}} = \xi _1,
\label{eq:Kozai1}
\end{equation}
and
\begin{equation}
3{\cos ^2}{i_{\rm tot}}(1 + 4e_1^2) - 1 - 9e_1^2 = \xi _2.
\label{eq:Kozai2}
\end{equation}

The constants $\xi _1$ and $\xi _2$ are determined by conditions of planets in any time through
\begin{equation}
{\xi _1}=G_1^2 + 2{G_1}{G_2}\cos {i_{\rm tot}},
\end{equation}
and
\begin{equation}
{\xi _2}=(1 + \frac{3}{2}e_1^2)(3{\cos ^2}{i_{\rm tot}} - 1) + \frac{{15}}{2}e_1^2{\sin ^2}{i_{\rm tot}}\cos (2{\omega _1}).
\end{equation}
Where,
\begin{equation}
\begin{array}{l}
{L_1} = \frac{{{M_{*}}{m_1}}}{{M_{*} + {m_1}}}\sqrt {G(M_{*} + {m_1}){a_1}}, \\
\\
{L_2} = \frac{{{m_2}({M_{*}} + {m_1})}}{{M_{*} + {m_1} + {m_2}}}\sqrt {G(M_{*} + {m_1} + {m_2}){a_2}},\\
\\
{G_1} = {L_1}\sqrt {1 - e_1^2}, \\
\\
{G_2} = {L_2}\sqrt {1 - e_2^2}.
\end{array}
\end{equation}

$e_2$ is the eccentricity of outer planet,which is constant. $i_{\rm tot}$ is the mutual inclination of two planets. when ${\omega _1} = \frac{\pi }{2}$, the eccentricity of inner planet reach the maximum value, while the inclination reach the minimum value in periodic variations. Solve Eq.(\ref{eq:Kozai1}) and (\ref{eq:Kozai2}), we can get the maximum value of $e_{1 }$, $e_{\rm max }$, in Kozai cycles, whatever libration or circulation. Here, if $a_1(1 - {e_{\rm max }}) \le 0.05$ au, we think a HJ can be formed through Kozai-Lidov mechanism. In the quadrupole approximation, the inner planet undergoing Kozai-Lidov mechanism can't flip. In the octupole approximation, some inner planet with extreme initial conditions can flip \citep{Naoz2011}. However there isn't an explicit analytical criterion to predict orbital flip.  In this article, we adopt quadrupole approximation to check Kozai-Lidov mechanism, and regard the inclination of planets in the end of integration as its' perpetual orbital orientation in continued evolution. 

4.  Secular chaos (SC).

Secular chaos is a mechanism to get equipartition of kinetic energy among the secular degrees of freedom \citep{Laskar1989, Laskar1997}. In planetary systems with two or more well-spaced, eccentric and inclined planets, the angular momentum deficit (AMD) is defined by
\begin{equation}
{\rm AMD} \equiv \sum\limits_{n = 1}^N {{\Lambda _n}(1 - \sqrt {1 - e_n^2} \cos {i_n})},
\label{eq:AMD}
\end{equation}
with
\begin{equation}
{\Lambda _n} = \frac{{m{M_{*}}}}{{m + {M_{*}}}}\sqrt {G({M_{*}} + m){a_n}},
\end{equation}
where, $i_n$ is its inclination relative to the invariable plane (normal to total angular momentum).

In the multi-planet systems under secular interaction among planets, the formation of HJs requires AMD of the whole system
\begin{equation}
{\rm AMD} \ge {\Lambda _1}\left[ {1 - {{(\frac{{{a_1}}}{{0.1{\rm AU}}})}^{{{ - 1} \mathord{\left/
 {\vphantom {{ - 1} 2}} \right.
 \kern-\nulldelimiterspace} 2}}}\cos {i_f}} \right],
\label{eq:AMD1}
\end{equation}
where, $i_f$ is the final inclination of HJ in invariant plane reference frame, and $a_1$ is the semi-major axes of inner planet which keeps almost constant before orbital circularization \citep{Wu2011}. We should note the criterion of secular chaos in Eq.(\ref{eq:AMD1}) is a necessary, but not sufficient, condition for producing HJs.  For example, some systems in Kozai-Lidov mechanism can also satisfy Eq.(\ref{eq:AMD1}), but the pericenter of inner planet $a(1 - {e_{\rm max }}) \ge 0.05$ au. 

  Because of the same dynamical equations, CHEM and Kozai in our manuscript belong to (eccentric) Kozai-Lidov mechanism. According to \citet{Li2014chaos} and \citet{Naoz2016}, the eccentric Kozai-Lidov effect shows as ``coplanar high-eccentricity migration" while octupole approximation dominates the disturbing function. \citet{Li2014chaos} pointed there are three scenarios that inner test particle may flip or obtain extreme eccentricities. The scenario ``when the inner orbit is eccentric and coplanar” is the High-e-Low-i situation (CHEM in this paper). ``When the inner orbit is circular and with high inclination” is the Low-e-High-i situation (Kozai in this paper). The scenario ``when the inner orbit is moderately eccentric and with very high inclination” is quite different with our two new mechanisms where the inner orbits flip with high eccentricities. In the general three body problem, more complicated situations may appear comparing with inner restricted three body problem \citep{Li2014chaos}. In this paper, we distinguish the different situations with specific dynamical characteristics in order to understand general three body problem more clearly.

From the descriptions of four criteria above, we have known that three situations, planet-planet scattering if only one planet remained, Kozai-Lidov mechanism in quadrupole approximation and Coplanar high-eccentricity migration can be accurately classified through orbital elements of planets in the end of integrations. However, classifying other situations needs the evolutions of planets. 

\subsection{Procedure to Distinguish Different Mechanisms}\label{sec:steps}

In order to tell different mechanisms during the evolutions of planetary systems, we check the orbital parameters of planets survived after ${10^7}$$T_{\rm in}$ as follows.  

1.  Check the number of planets survived. If there is a single planet remained, and ${a_1}(1 - {e_1}) \le 0.05$ au, PPS is the only way to produce it.  If there is more than one planet remained, turn into next step.

2.  If the mutual inclinations between inner and outer planets $ I_{mut} \le {20^ \circ }$, and the systems satisfy Eq.(\ref{eq:CHEM1}) or Eq.(\ref{eq:CHEM2}), the inner planet can become hot Jupiter through CHEM.

3. Put the orbital elements of planets into the Eq.(\ref{eq:Kozai1}) or Eq.(\ref{eq:Kozai2}), solve them simultaneously to obtain the maximum eccentricity of inner planet. If there is a meaningful solution, and it satisfies $a_1(1 - {e_{\rm max }}) \le 0.05$ au, we classify this mechanism to Kozai.

4.  We pick out the systems with ${a_1}(1 - {e_1}) \le 0.05$ au in the end of integrations as well as the systems satisfying Eq.(\ref{eq:AMD1}),  ${i_f}=0^ \circ$ from the cases remained. The evolutions of systems are needed to make clear the mechanisms they went through. In the systems satisfying Eq.(\ref{eq:AMD1}), $i_f=0^ \circ$, we export the orbital elements of planets survived with interval ${10^3}$$T_{\rm in}$. For the inner planet which satisfies $a_1(1 - {e_{\rm max }}) \le 0.05$ au more than 20 times, we simply think it can be circularized by tidal effects \citep{Nagasawa2008}.  We plot the phase diagram of $e_1 \sim \omega_1 $ and $e_1 \sim \Delta \varpi $ to affirm which mechanism they belong to，where $\Delta \varpi = \varpi_1 - \varpi_2$.

4A. If the phase diagram of $e_1 \sim \omega_1 $ of inner planet have the characteristics of Kozai, we classify these situations as Kozai.

4B. If the phase diagram of $e_1 \sim \Delta \varpi $ of inner planet have the characteristics of CHEM, we classify these situations as CHEM.

4C. If semi-major axes of planets after dynamical instability remain almost constant, the phase diagram of $e_1 \sim \omega_1 $ and $e_1 \sim \Delta \varpi $ doesn't have the characteristics of Kozai and CHEM, the evolutions of eccentricities and inclinations of planets are non-periodic, we classify these situations as SC.

4D. The other situations are categorized into PPS.

In next section, we will give the simulation results about efficiencies of HJs formation in relation to several parameters and different dynamical mechanisms.

\section{NUMERICAL RESULTS}\label{sec:szjg}
\subsection{Two New Mechanisms to Form Hot Jupiters }\label{sec:TNM}

During the process of classifications on different high-eccentricity mechanisms to form HJs, we find other two new mechanisms to form high inclined HJs, even retrograde ones.

\begin{figure}[htbp]
  \centering
  \includegraphics[width= 1\textwidth]{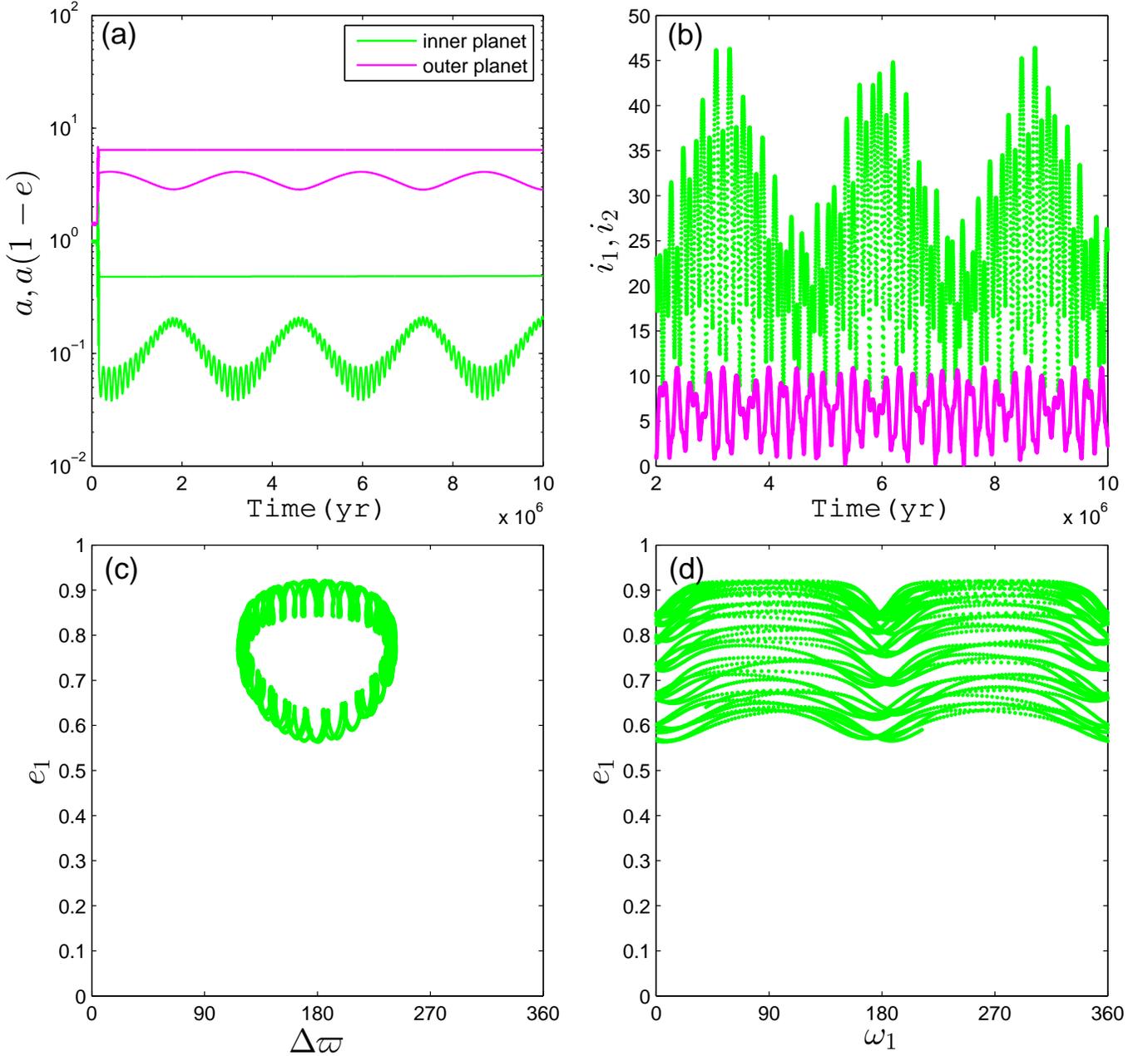}\\
  \caption{The new mechanism E1, in which CHEM modulate the eccentricity of inner planet in a long period, while Kozai modulate in a short period, can produce highly inclined  HJs. The evolutions of three planet system with initial inner planet located at 1 au, all orbits initially circular, inclinations $ i _{1}= 1.13^ \circ$, $ i _{2}= 1.63^ \circ$, $ i _{3}= 0.51^ \circ$, and space separation $ k = 4.105$. After about $2 \times {10^6}$ yr, the survived two planets entry into E1 mechanism. We just show the evolutions of the planets survived for other planets are ejected out or hit into the host. We can reproduce this mechanism through the orbital elements of inner planet $ a _{1}=0.48697$,  $ e _{1}=0.662205$,  $ i _{1}=11.13053^ \circ$, $ \omega_1=271.9666^ \circ$  ${\Omega _1}=87.7832^ \circ$, ${M_1}=277.4253^ \circ$,
 and the orbital elements of outer planet $a _{2}=6.41010$,  $e _{2}=0.546336$,  $i _{2}=7.634003^ \circ$, $ \omega_2=351.4251^ \circ$,  ${\Omega _2}=221.8703^ \circ$, ${M_2}=106.9747^ \circ$. The remained planets didn't collide with each other, so their mass didn't change.  
(a) shows the whole evolutions of semi-major axes and pericenter of planets from dynamical instability until entry into E2 mechanism. (b) shows the inclination of planets after $2 \times {10^6}$ yr. (c) gives the phase diagram of $e_1 \sim \Delta \varpi $ of inner planet after $2 \times {10^6}$ yr. (d) gives the phase diagram of $e_1 \sim \omega _1 $ of inner planet after $2 \times {10^6}$ yr.}\label{fig:CKP}
\end{figure}

\begin{figure}[htbp]
  \centering
  \includegraphics[width= 1\textwidth]{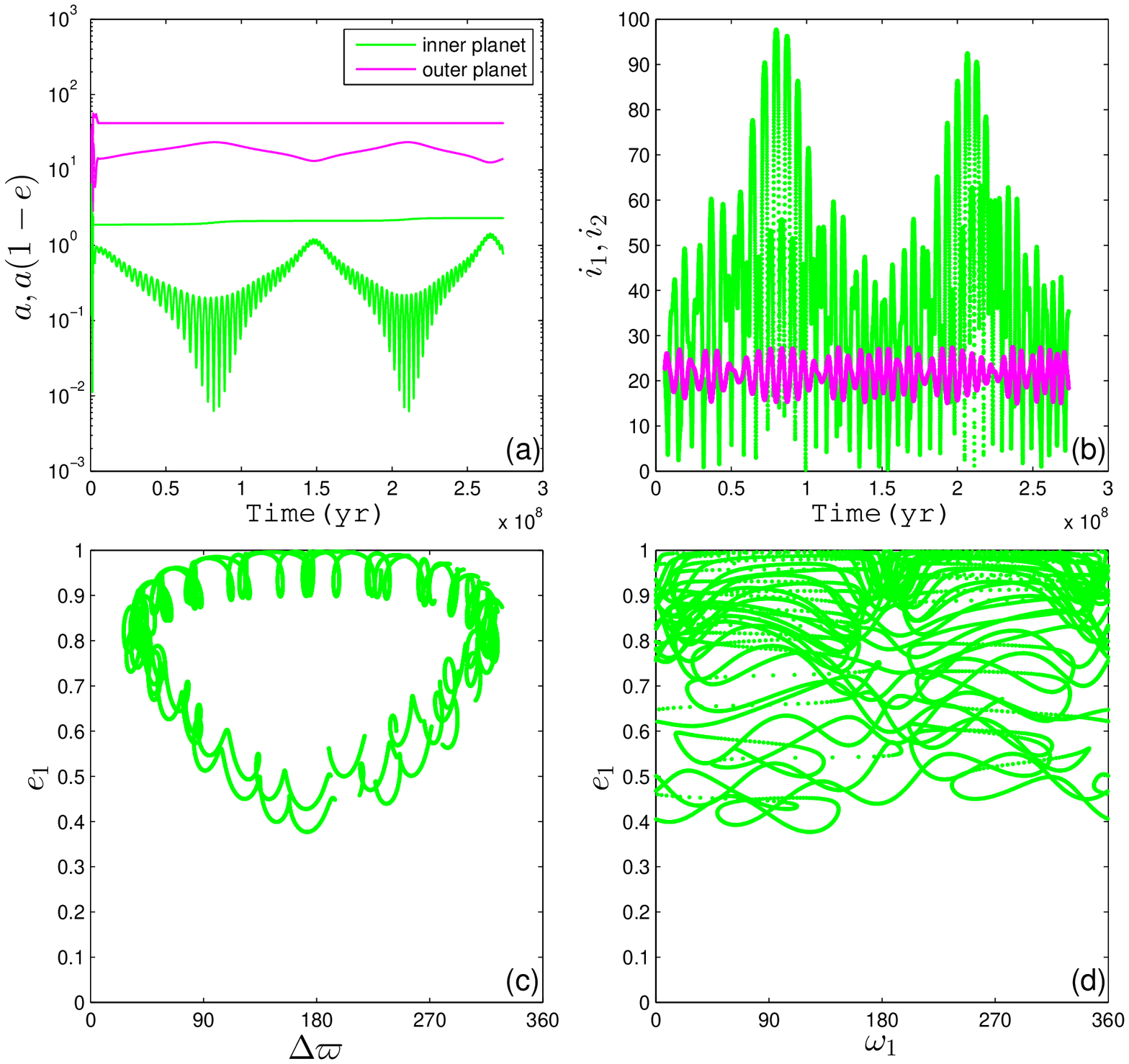}\\
  \caption{Retrograde inner planet can form through E1 mechanism. The evolutions of four planet system with initial inner planet located at 5 au, all orbits initially circular, inclinations $ i _{1}= 0.55^ \circ$ ,$ i _{2}= 0.88^ \circ$, $ i _{3}= 1.29^ \circ$, $ i _{4}= 0.64^ \circ$, and space separation $ k = 3.241$. After about $6 \times {10^6}$ yr, the survived two planets entry into E1 mechanism. We just show the evolutions of the planets survived for other planets are ejected out or hit into the host. We can reproduce this mechanism through the orbital elements of inner planet $ a _{1}=2.29327$,  $ e _{1}=0.527264$,  $ i _{1}=3.486438^ \circ$, $ \omega_1=284.2394^ \circ$  ${\Omega _1}=351.8428^ \circ$, ${M_1}=334.4163^ \circ$, 
and the orbital elements of outer planet $a _{2}=41.71439$,  $e _{2}=0.686280$,  $i _{2}=26.09256^ \circ$, $ \omega_2=199.10571^ \circ$,  ${\Omega _2}=330.0275^ \circ$, ${M_2}=60.6561^ \circ$.
 The remained planets didn't collide with other planets, so their mass didn't change.  
(a) shows the whole evolutions of semi-major axes and pericenter of planets from dynamical instability until entry into E2 mechanism. (b) shows the inclinations of planets after $6 \times {10^6}$ yr. (c) gives the phase diagram of $e_1 \sim \Delta \varpi $ of inner planet after $6 \times {10^6}$ yr. (d) gives the phase diagram of $e_1 \sim \omega _1 $ of inner planet after $6 \times {10^6}$ yr.}\label{fig:CKR}
\end{figure}

First, a mechanism, marked as E1, have the characteristics of both CHEM and Kozai, which greatly enlarge the inclination of inner planet in pure CHEM.  And, in this mechanism, when the inner planet move closest to the host, its inclination reach the maximum, different from Kozai-Lidov mechanism. For example, in Fig. \ref{fig:CKP}, after dynamical instability of an initial three-planet system, the remained two planets enter into E1 mechanism. The $ \Delta \varpi $ of inner planet librate and  $\omega_1 $ circulate. Interestingly, this mechanism overcomes the difficulty of CHEM to produce high inclined HJs. The inclinations of HJs formed through CHEM are mostly $\le {30^ \circ }$ \citep{Petrovich2015a}. However, the inclinations of HJs obtained in E1 mechanisms can be as high as $> {30^ \circ }$, even retrograde. The inclination and eccentricity of prograde inner planet vary in opposite direction in quadrupole Kozai-Lidov mechanism \citep{Lithwick2011}. During the flip of planets in Kozai-Lidov mechanism, when the eccentricities reach the maximum, the inclinations are about $90^ \circ$. However, we find a positive correlation between inclination and eccentricity of inner planet in E1 mechanism. This means, the orbits of inner planets in E1 mechanism are more easily circularized by tidal effects during the states of maximal inclinations in periodic variations, which may result in HJs with high inclinations. While, the orbits of prograde inner planets in Kozai mechanism are more easily circularized during the states of minimal inclinations in periodic variations, and the flipping planets in Kozai mechanism are more possibly circularized during inclinations around $90^ \circ$ \citep{Naoz2011}. One sample of retrograde orbit produced through E1 mechanism is shown in Fig. \ref{fig:CKR}. 

\begin{figure}[htbp]
  \centering
  \includegraphics[width= 1\textwidth]{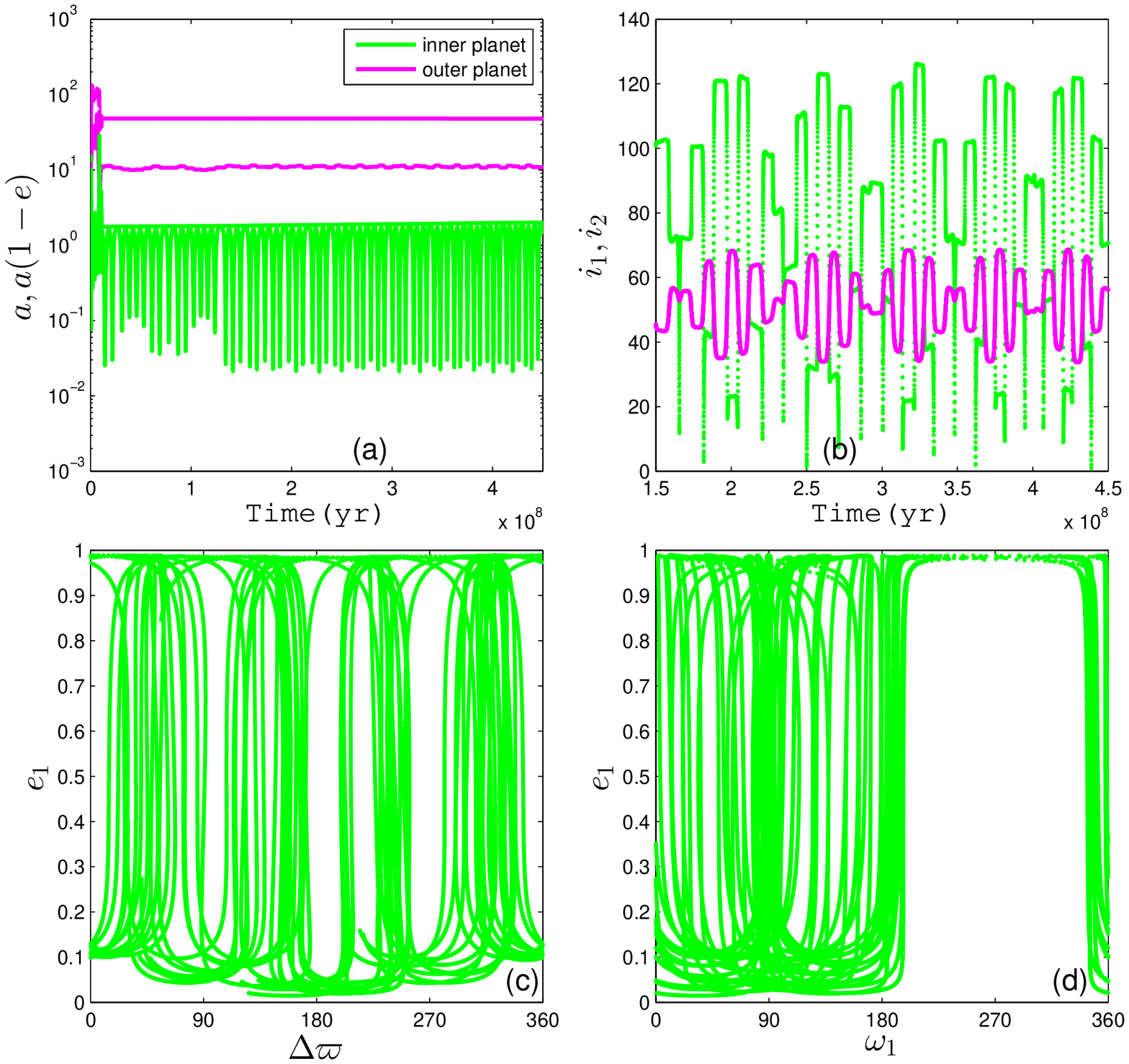}\\
  \caption{ The evolutions of a five-planet system with the inner planet initially located at 5 au. All orbits are initially circular with inclinations  $ i _{1}= 1.38^ \circ$ ,$ i _{2}= 1.04^ \circ$, $ i _{3}= 1.39^ \circ$, $ i _{4}= 0.84^ \circ$, $ i _{5}= 0.84^ \circ$, and space separation $ k = 3.498$.  After about $1.5 \times {10^8}$ yr, the survived two planets entry into E2 mechanism. We just show the evolutions of the planets survived for other planets are ejected out or hit into the host. We can reproduce this mechanism through the orbital elements of inner planet $ a _{1}=1.826147$,  $ e _{1}=0.053076$,  $ i _{1}=72.12216^ \circ$, $ \omega_1=164.6919^ \circ$  ${\Omega _1}=118.8519^ \circ$, ${M_1}=200.9330^ \circ$, 
and the orbital elements of outer planet $a _{2}=48.04923$,  $e _{2}=0.772293$,  $i _{2}=55.79706^ \circ$, $ \omega_2=318.3559^ \circ$,  ${\Omega _2}=16.1538^ \circ$, ${M_2}=188.0218^ \circ$. (a) shows the whole evolutions of semi-major axes and pericenter of planets from dynamical instability until entry into E2 mechanism. (b) shows the inclination of planets after$1.5 \times {10^8}$yr. (c) gives the phase diagram of $e_1 \sim \Delta \varpi $ after $1.5 \times {10^8}$yr. (d) gives the phase diagram of $e_1 \sim \omega _1 $ after $1.5 \times {10^8}$yr.}\label{fig:unknown}
\end{figure}

Second, in another new mechanism, marked as E2,  the eccentricity of inner planet changing  periodically,  it's orbit can flip directly between prograde and retrograde, for example in Fig. \ref{fig:unknown}.  The phase diagram of $e_1 \sim \omega_1 $ doesn't show the characteristic of Kozai. And the phase diagram of $e_1 \sim \Delta \varpi $ doesn't show the characteristic of CHEM.  Kozai-Lidov mechanism in octupole approximation can also lead to flipping inner orbits. However, the variation of inclination is gradual, unlike a sudden change in E2.  In CHEM, when the inner orbit flip from around $0^ \circ$ dirctly to around $180^ \circ$, it's eccentricity usually has $1 - {e_1} \lesssim {10^{ - 3}} - {10^{ - 4}}$ \citep{Li2014}. While, in E2 mechanism, the eccentricity of inner planet $1 - {e_1} > {10^{ - 2}}$, so the orbital circularization by tidal effect will not be disturbed \citep{Guillochon2011}. E2 mechanism shouldn't belong to the flipping situation in Kozai or CHEM, but a new mechanism.  Once the eccentricity reaches the maximum, the inclination of inner planet would vary, which is well coupled with the inclination of outer planet.

{/bf The new mechanisms are calculated with Bulirsch-Stoer method which is the most accurate integrator in Mercury code. We reproduced the mechanism in Fig.\ref{fig:CKP}- \ref{fig:unknown} perfectly with BS2 method, well a little distortion with MVS and RADAU method. }
For the aim of this article is to study HJs formation from high-eccentricity mechanisms, so detailed researches on initial conditions and theories of above two new mechanisms will be given in another paper.

\subsection{Efficiencies of Different Mechanisms to Form HJ candidates }\label{sec:effm}

After classifications on six mechanisms in Section\ref {sec:condition} and \ref{sec:TNM} by steps in Section \ref{sec:steps}, we obtain the respective numbers of cases undergoing or underwent corresponding mechanisms. For the group with initial inner semi-major axes at 1 au,  PPS : Kozai : SC : CHEM : E1: E2 = $303 : 229 : 18 : 54 : 93 : 15$; For the group with initial inner semi-major axes at 5 AU, PPS : Kozai : SC : CHEM : E1: E2 = $ 79 : 172 : 3 : 28 : 84 : 16$.   
Among the six high-eccentricity mechanisms, PPS, Kozai and E1 are more effective, while, E2 and SC are less valid in our simulations.

Formation of HJs through strong chaos requires high-eccentricities and/or inclinations in two-planet systems \citep{Naoz2011}. In the systems consisting of more than two planets, HJs forming through secular chaos are relatively easy \citep{Wu2011}. However, the number of planets survived after dynamical instability is usually less than three. For examples, in the systems which can produce HJ candidates, when $N = 4$, the ratio of systems with more than two planets survived is about $2.5\% $, while for $N = 5$, about $5.5\% $. The two values almost keep constant with the change of initial inner semi-major axes. So, sufficient conditions for HJs formation by secular chaos are hardly obtained after dynamical instability of multi-planet systems. But, if dynamical instability of multi-planet systems doesn't occur, the sufficient AMD to produce HJs aren't achievable within ${10^7}$$T_{\rm in}$ in our simulations.. 

The total case numbers of HJ candidates formation decrease with the increasing of ${a_{\rm in}}$. Among all the mechanisms, the possibility of HJ candidates formed by PPS decreases from $21\%$ in the group of ${a_{\rm in}}= 1$ au to $43\%$ when ${a_{\rm in}}= 5$ au. Because PPS includes the HJ candidates formed by transient tidal circularization before the systems turn to long term interactions.  This part of HJ candidates are closely related to the ${a_{\rm in}}$, for the inner planets can reach tidal boundaries of hosts without a high eccentricities during fierce planet-planet interactions. 

Except PPS, all the other mechanisms are long term planet-planet interactions. Especially, Kozai takes great effect on the formation of HJ candidates in both pure Kozai-Lidov mechanism and E1 mechanism where Kozai coupling with CHEM. The total parts of Kozai participating in are $45\%$ of HJ candidates formed in the group ${a_{\rm in}}= 1$ au and $67\%$ for ${a_{\rm in}}= 5$ au. 

\subsection{Relations between Efficiencies of HJ candidates Formation and Initial Planet Numbers}\label{sec:effN}

We find the situations of HJ candidates formed in initial two-planet systems and those consist of more than two planets are strikingly different.  In systems consisting of two planets with ${a_{\rm in}}= 1$ au, only if the initial space separation between two planets is quite small ( $k < 3.5$ ), HJ candidates can form through PPS  mechanism. This is because a planet pair with mutual separation $k \ge 3.5R_{H}$, they are Hill stable \citep{Gladman1993}. And it's efficiency is only $1\% $, which reduce to 0 when ${a_{\rm in}}= 5$ au even with $k < 3.5$. If initial planet number $N \ge 3$, the multi-planet systems with the same ${a_{\rm {in}}}$ have comparable efficiencies of HJ candidates formation, $7\%  \sim 9\% $ for ${a_{\rm {in}}}= 1$ au and $4\%  \sim 5\% $ for ${a_{\rm in}}= 5$ au.

Kinetic energy of planets can be converted to heat energy during collisions among planets. However, in our simulations, planetary collisions never occur.  After all the other planets were ejected or hit into the host, the final semi-major axis of the planet ${a_f}$ can be obtained approximately by conservation of energy. Neglecting the mass of planets which hit into the star,
\begin{equation}
{a_f} \approx  - \frac{{GM_{*}m}}{{2K}}.
\end{equation}
Where, $K$ is the orbital energy of the planet survived, which is approximately equal to the linear summation of initial orbital energy of planets,
\begin{equation}
K \approx \sum\limits_{n = 0}^N { - \frac{{G M_{*}m}}{{2{a_n}}}}, 
\end{equation}
In our simulations, the planets have equal Jupiter mass. The initial semi-major axes of planets rank in geometric series distribution, with the ratio $q$,
\begin{equation}
q = \frac{{{a_{n + 1}}}}{{{a_n}}} = {\frac{{1 + \frac{k}{2}\left( {\frac{{2\mu }}{3}} \right)}}{{1 - \frac{k}{2}{{\left( {\frac{{2\mu }}{3}} \right)}^{{1 \mathord{\left/
 {\vphantom {1 3}} \right.
 \kern-\nulldelimiterspace} 3}}}}}^{{1 \mathord{\left/
 {\vphantom {1 3}} \right.
 \kern-\nulldelimiterspace} 3}}}.
\end{equation}
So, the final semi-major axis of the planet survived,
\begin{equation}
{a_f} \approx \frac{{\left( {1 - \frac{1}{q}} \right)}}{{1 - \frac{1}{{{q^N}}}}}{a_{in}}.\label{eq:af}
\end{equation}
From the equation above, we find ${a_f}$ decreases with closer ${a_{in}}$, smaller space separation, and more planets.  ${a_f}$ has the limit value $ \left( {1 - \frac{1}{q}} \right){a_{in}}$, if $N \to \infty $. That means the influence of planet number on ${a_f}$ will be weakened gradually. 

\begin{table}[htbp]
\centering
\begin{tabular}{|c|c|c|c|c|c|c|c|c|}
\hline
N&PPS & CHEM & Kozai &SC & E1 & E2 & Total &NC \\
\hline
2 & 11  &  0          & 0        & 0   & 0   & 0 & 11 ($1\%$)  & 1100  \\
\hline
3 & 28  & 33        & 74       & 1   & 37  & 2 & 175 ($7.6\%$)  & 2300  \\
\hline
4 & 91  & 8          & 80       & 8   & 29  & 4 & 220 ($6.9\%$)  & 3200  \\
\hline
5 & 173 & 13      & 77       & 9   & 24  & 9 & 305 ($9.0\%$)  & 3400  \\
\hline
\end{tabular}
\caption{  The table gives the numbers of HJ candidates formed through different high-eccentricity mechanisms for different initial planet number in the group of ${a_{\rm in}}= 1$ au.  N is the initial planet number, NC is the number of cases which will become dynamical instability within ${10^7}$$T_{\rm in}$.  The percentages in brackets are the proportions of cases which can form HJ candidates among respective NC cases. }\label{tab:NHJs}
\end{table}

From Table \ref{tab:NHJs}, we find there is a distinct positive correlation between the efficiencies of PPS forming HJ candidates and initial planet numbers. With the increasing of planet numbers, perturbations on inner planets will be larger. During the mutual interactions of planets, the inner planets in the third kind of PPS are more easily to reach the tidal boundary of the central stars. While the other long term mechanisms don't have high requirement on the intensity of mutual scattering to obtain large enough eccentricities. SC can hardly happen in the cases with $N = 3$. In these cases, after dynamical instability, one or two planets are remained. However, HJs can hardly form in two-planet systems \citep{Wu2011}. When $N \ge 4$, the occurrence of SC to form HJ candidates isn't sensitive to $N$ any more. 

\subsection{Relations between Efficiencies of HJ candidates Formation and Space Separations}\label{sec:effk}

All of the previous studies on HJs formation were based on multi-planet systems with several groups of fixed space separtions \citep{Marzari2002,Chatterjee2008,Nagasawa2008,Nagasawa2011,Beauge2012}.  From researches on dynamical instability of multi-planet systems \citep{Zhou2007}, there is a strong positive correlation between dynamical instability timescales of systems and space separations between adjacent planets. And from Eq. (\ref{eq:af}), the final semi-major axis of planet survived will decrease with smaller space separation. Both of the intensity of mutual scattering and final semi-major axes of inner planets survived can exert influences on the formation of HJs.

\begin{figure}[htbp]
  \centering
  \includegraphics[width= 1\textwidth]{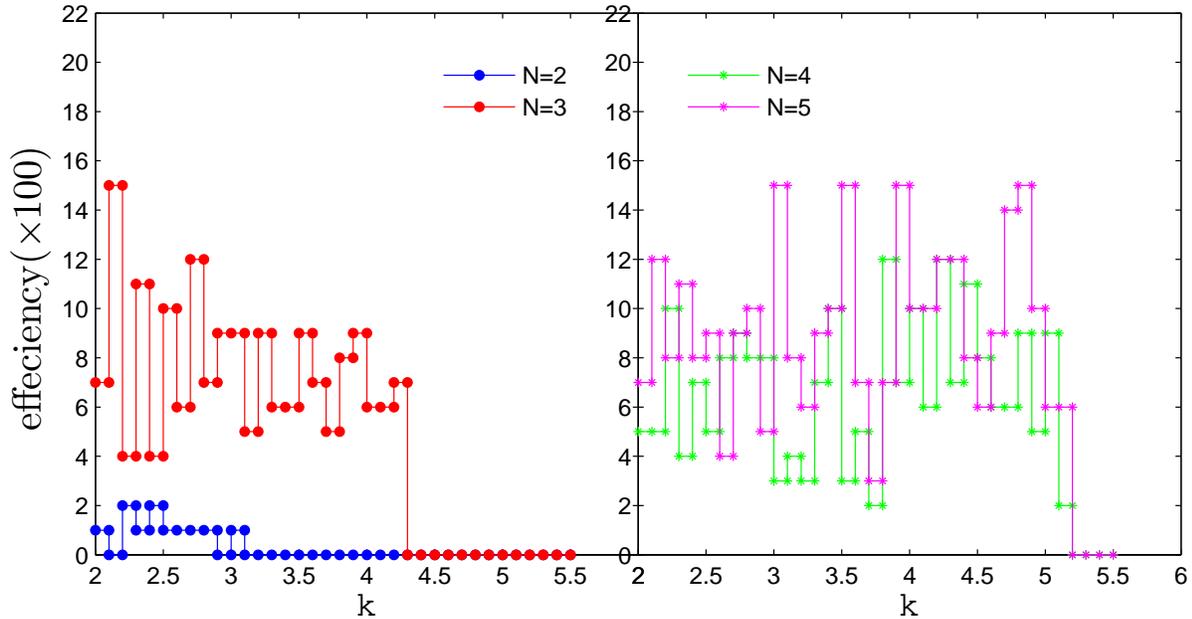}\\
  \caption{ The efficiencies of HJ candidates formation among 100 cases with space separation over the interval $\Delta k = 0.1$ within ${10^7}$ $T_{\rm in}$  while $a_{in} = 1$AU are shown in different colors for specific number of planet $N = 2 \sim 5$. }\label{fig:EK2}
\end{figure}

\begin{figure}[htbp]
  \centering
  \includegraphics[width= 1\textwidth]{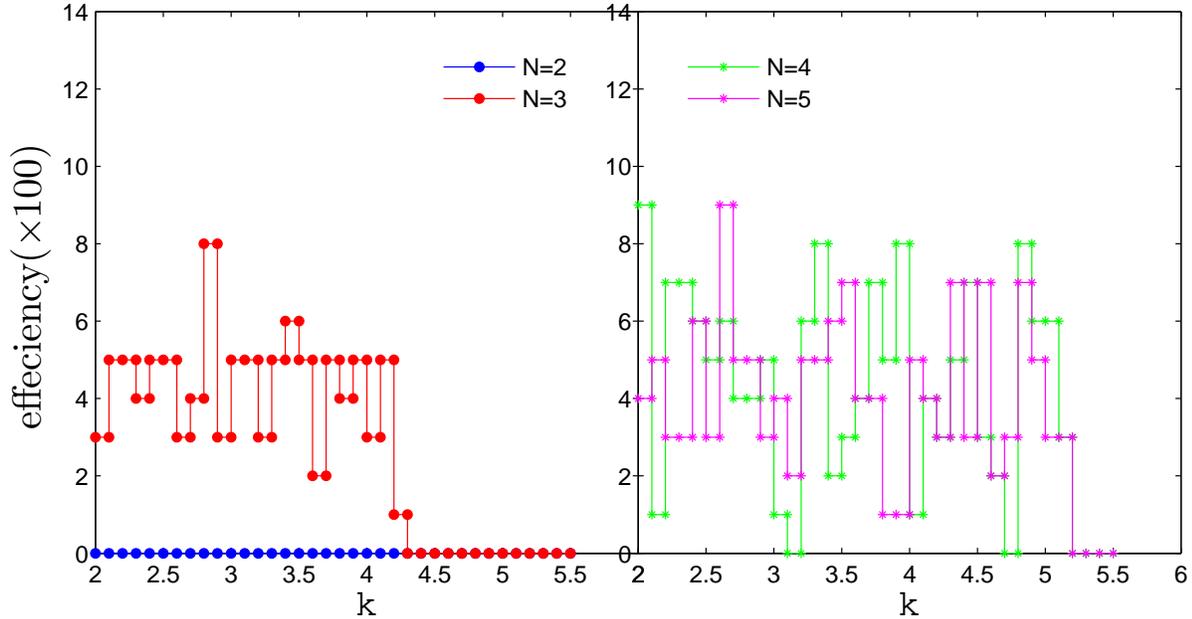}\\
  \caption{The efficiencies of HJ candidates formation among 100 cases with space separation over the interval $\Delta k = 0.1$ within ${10^7}$ $T_{\rm in}$  while $a_{in} = 5$AU are shown in different colors for specific number of planet $N = 2 \sim 5$. }\label{fig:EK4}
\end{figure}

First, the dynamical instability of multi-planet systems within ${10^7}$$T_{\rm in}$ determines the ranges of $k$ where HJs can form through high-eccentricity mechanisms. Fig. \ref{fig:EK2} and Fig. \ref{fig:EK4} show the efficiencies of HJ candidates formation related to $k$ in two groups of $a_{in}$.  The range of space separations where HJs maybe form within ${10^7}$$T_{\rm in}$ are as follows:  ${\rm N}=2$, $k \le 3.1$; ${\rm N}=3$, $ k \le 4.3$; ${\rm N}=4,5$, $ k \le 5.2$. The significances of Fig. \ref{fig:EK2} and Fig. \ref{fig:EK4} are that they offer us the efficiencies of HJ candidates formation about different $k$ among 100 cases for respective initial planet numbers.  

Second, from Eq.(\ref{eq:af}), we know, with closer distance between planets, the final semi-major axes of planets remained will be smaller. This reduces the requirements on eccentricities of inner planet needed to reach tidal boundary. However, as shown in Fig. \ref{fig:EK2} and Fig. \ref{fig:EK4}, we find no explicitly decreasing trends in efficiencies of HJ candidates formation with $k$.
The locations of first-order mean motion resonances $2 : 1$, $3 : 2$ and $4 : 3$ corresponds to $k \approx 5.20$, $k \approx 4.58$ and $k \approx 3.27$.  For planetary orbital angles chosen randomly, most of cases with space separations around mean motion resonances are unstable. So efficiencies of HJ candidates formation drop dramatically around $k \approx 4.58$ and $k \approx 3.27$. 

\begin{figure}[htbp]
  \centering
  \includegraphics[width= 1\textwidth]{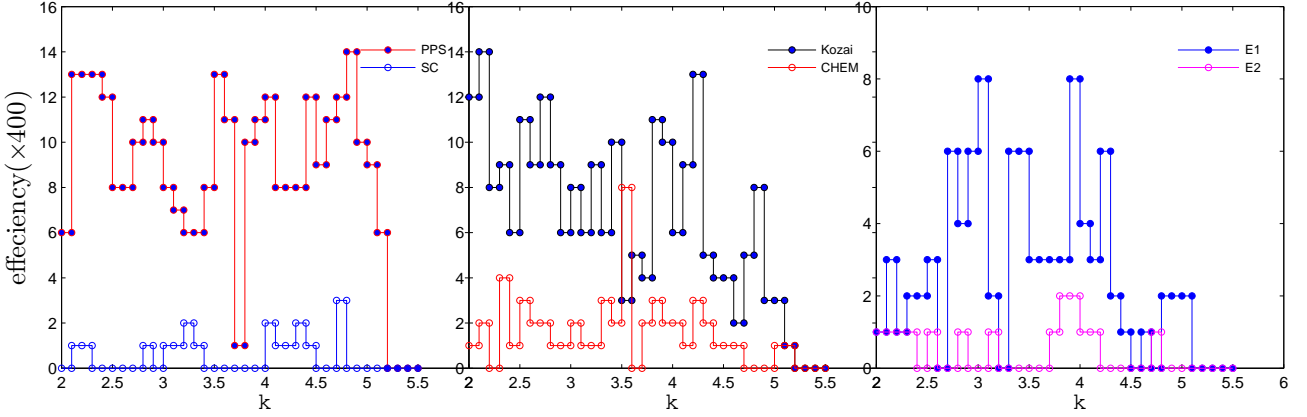}\\
  \caption{The efficiencies to form HJ candidates through different mechanisms, PPS, Kozai-Lidov mechanism, coplanar high-eccentricity migration, secular chaos and the other two new mechanisms we found, with specific space separation between adjacent planets while $a_{in} = 1$ au.}\label{fig:MEK2}
\end{figure}

\begin{figure}[htbp]
  \centering
  \includegraphics[width= 1\textwidth]{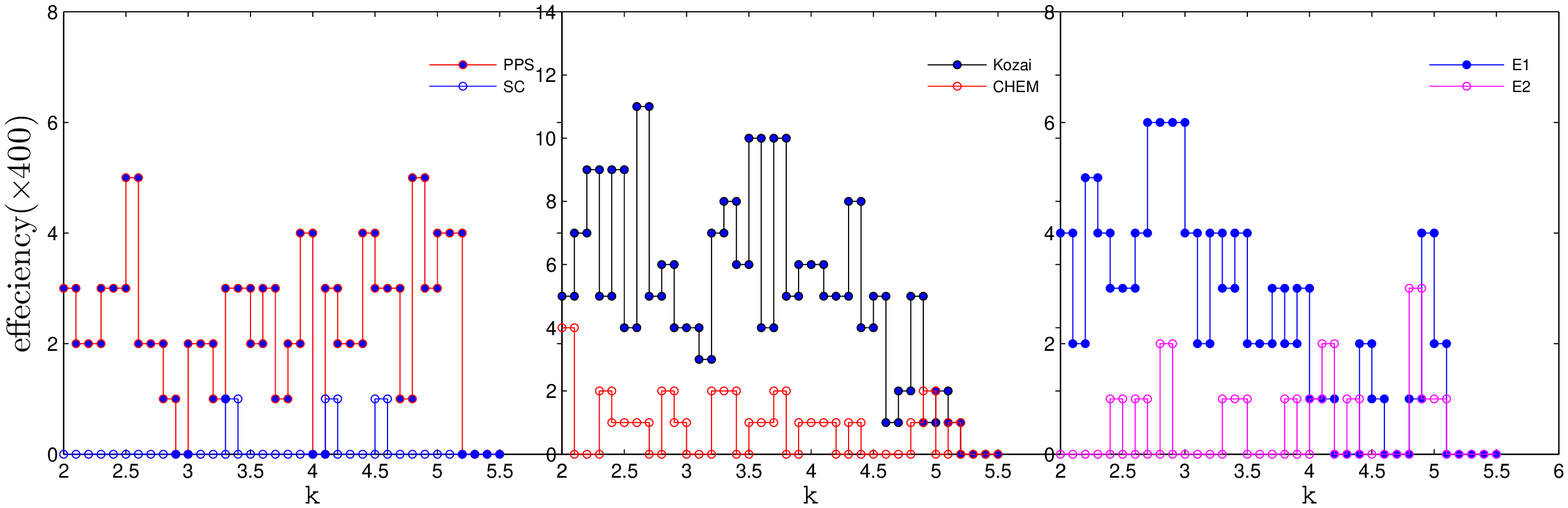}\\
  \caption{The efficiencies to form HJ candidates through different mechanisms, PPS, Kozai-Lidov mechanism, coplanar high-eccentricity migration, secular chaos and the other two new mechanisms we found, with specific space separation between adjacent planets while $a_{in} = 5$ au.}\label{fig:MEK4}
\end{figure}

Third, the efficiencies of six high-eccentricity mechanisms in our article to form HJ candidates with varied space separations are shown in Fig. \ref{fig:MEK2} and Fig. \ref{fig:MEK4}. Interestingly, the cases of secular chaos gather around mean motion resonances, while other mechanisms fall to the valley floors. A deceptive phenomenon appear in PPS mechanism, the efficiencies of which increase near $k = 5.2$. For some systems with space separation near $k = 5.2$ just become dynamical instability, and a planet reach the tidal boundary of the host in a short time. But these planets may be ejected out or hit into the host in subsequent revolution. We continue integrating these systems, and find they enter into final stable states with one or two planets remained after ${10^8}$ $T_{\rm in}$.  So, the efficiency of PPS around the critical $k$ value are overstated. We know that multi-planet systems are intrinsically chaotic. No matter how longer the integrations are, this kind of situations will occur around the corresponding critical $k$ of the integration time \citep{Zhou2007arxiv}. This part of HJ candidates produce a maximum error $6\% $ on the total cases in which can form HJ candidates.

\section{COMPARISON WITH OBSERVATIONS}\label{sec:observation}

\subsection{Proportion of Retrograde HJ candidates}\label{sec:nr}

In our simulations, except CHEM, all the other high-eccentricity mechanisms can produce retrograde HJs.  In order to distinguish different mechanisms of HJs formations, we didn't add tidal force to numerical simulations, so final inclinations of HJs after orbital circularization by tidal effects are unknown. But we can get the lower limit of proportion of retrograde HJ candidates by only considering the cases of inner planets keeping retrograde. Also we can get the upper limit of proportion of retrograde HJ candidates by considering additional cases of inner planets flipping between prograde and retrograde orbits. The two limits with different $a_{in}$ and each mechanism are listed in Table \ref{tab:ReHJs}.   
\begin{table}[htbp]
\centering
\begin{tabular}{|c|c|c|c|c|c|c|c|}
\hline
    &PPS & CHEM & Kozai &SC & E1 & E2 & Total \\
\hline
${a_{\rm in}}= 1$ au & $0.9\%$ ($43\%$) & $0\%$($0\%$) & $4\%$($10\%$)  & $0\%$ ($6\%$)& $0\%$ ($2\%$)  & $0\%$($53\%$) & $2\%$($23\%$)  \\
\hline
${a_{\rm in}}= 5$ au & $5.3\%$($75\%$) & $0\%$($0\%$) & $13.4\%$($43\%$) & $0\%$($0\%$) & $0\%$($12\%$) & $0\%$($88\%$) & $8\%$($40\%$) \\
\hline

\end{tabular}
\caption{The upper and lower limits of retrograde proportion among HJ candidates formed through different mechanisms. The lower limit of proportion of retrograde HJ candidates are obtained by only considering the cases of inner planets keeping retrograde. The upper limits (in brackets) are obtained by add up the cases of inner planets keeping retrograde and the cases of inner planets flipping between prograde and retrograde orbits. }\label{tab:ReHJs}
\end{table}

From Table \ref{tab:ReHJs}, we find the total proportion, both the upper and lower limits of retrograde HJ candidates with ${a_{\rm in}}= 5$ au are much higher than those of ${a_{\rm in}}= 1$ au.  The absolute numbers of retrograde HJ candidates are almost the same in the two groups. The main reason is that the absolute number of multi-planet systems can produce HJ candidates in the group of ${a_{\rm in}}= 1$ au is about twice as much as the group of ${a_{\rm in}}= 5$ au. The total angular momentum of planets survived is determined by the loss of angular momentum carried on the ejected planets. Due to the random escaping orientation of ejected planets, more ejected planets bring higher possibility to change the component of angular momentum perpendicular to the direction of total angular momentum, then, lead to higher inclinations of planets survived. The number of ejections in the group of ${a_{\rm in}}= 5$ au is $10\% $ more than the group of ${a_{\rm in}}= 1$ au.
Collisions between planet and the host don't change the total angular momentum almost.

In the multi-planet systems with initial two planets, all the HJ candidates formed are prograde for the lack of fierce interactions. For one or two planets survived after dynamical instability in most systems, with the initial planet number increasing, more planets will be ejected out.  So, the possibility of forming prograde HJ candidates increase with more initial planet number. 


\begin{table}[htbp]
\centering
\begin{tabular}{|c|c|c|c|c|}
\hline
    &N=2 & N=3 & N=4 &N=5  \\
\hline
${a_{\rm in}}= 1$ au & $0\%$ ($0\%$) & $1.7\%$($10\%$) & $1.8\%$($17.7\%$)  & $2.3\%$ ($34.8\%$)  \\
\hline
${a_{\rm in}}= 5$ au & $0\%$ ($0\%$)  & $9.1\%$($22\%$) & $6.2\%$($41.1\%$) & $9.6\%$($55.9\%$)  \\
\hline

\end{tabular}
\caption{The upper and lower limits of retrograde proportion among HJ candidates formed with respective initial planet numbers. The lower limit of proportion of retrograde HJ candidatesare obtained by only considering the cases of inner planets keeping retrograde. The upper limits (in brackets) are obtained by add up the cases of inner planets keeping retrograde and the cases of inner planets flipping between prograde and retrograde orbits. The values of upper limits are in brackets. }\label{tab:ReNHJs}
\end{table}

 We will make further studies through the companion of HJ candidates in next section. 

\subsection{Characteristics of Planetary Companions of HJ candidates}\label{sec:nr}

\begin{figure}[htbp]
  \centering
  \includegraphics[width= 1\textwidth]{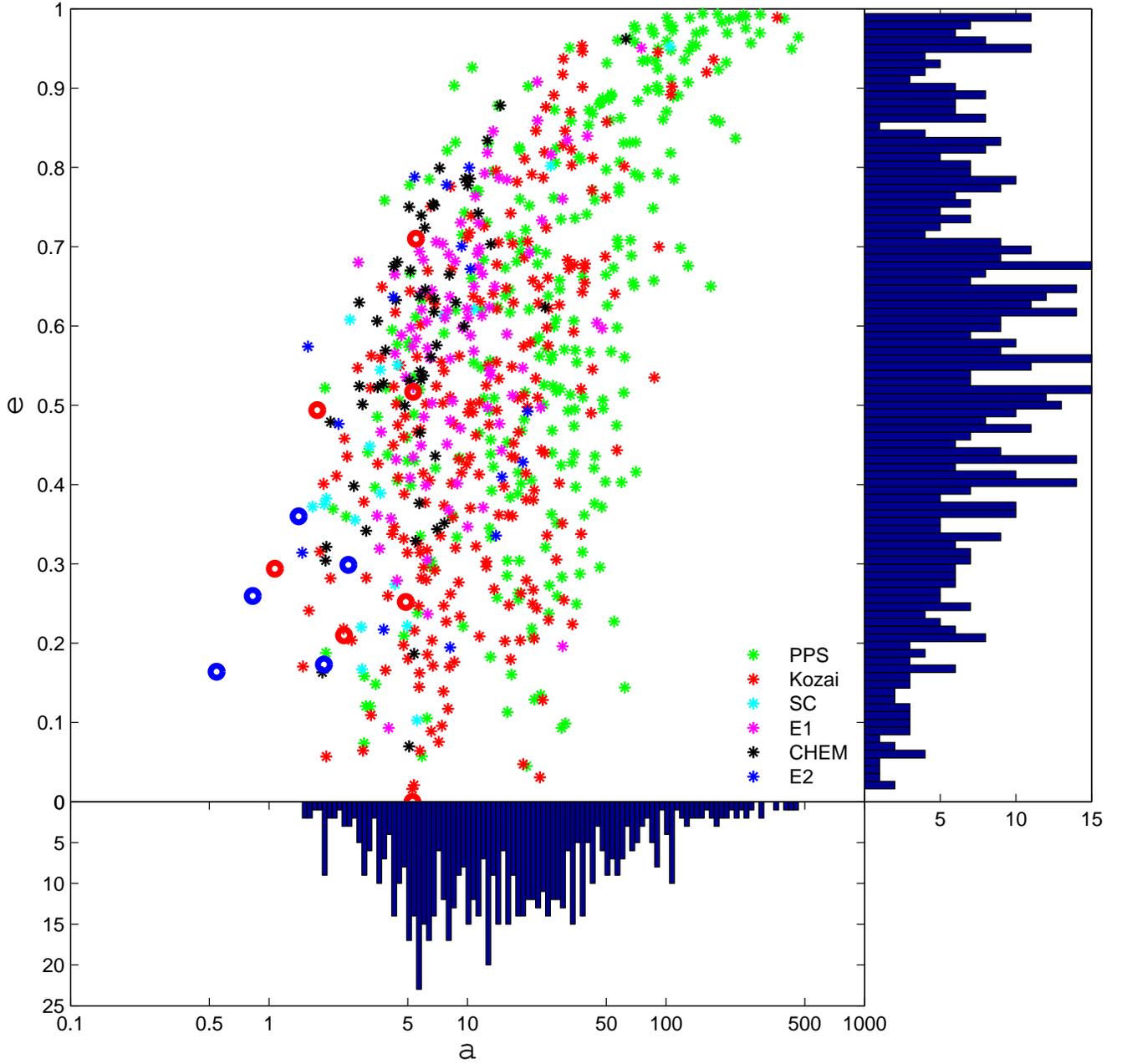}\\
  \caption{Scattering plot of semi-major axes and eccentricities of companions of HJ candidates in our simulations with ${a_{\rm in}}= 1$ au is shown in main diagram. We distinguish PPS, Kozai, CHEM, SC, E1, E2 using different colors.  Respective histograms of $a$, $e$ locate in bottom and right. The hollow blue circles represent the observed companions of HJs in ups And, Wasp-47 and HIP 14810, while the hollow red circles represent the others in Table \ref{tab:HJC}. }\label{fig:ae2}
\end{figure}

\begin{figure}[htbp]
  \centering
  \includegraphics[width= 1\textwidth]{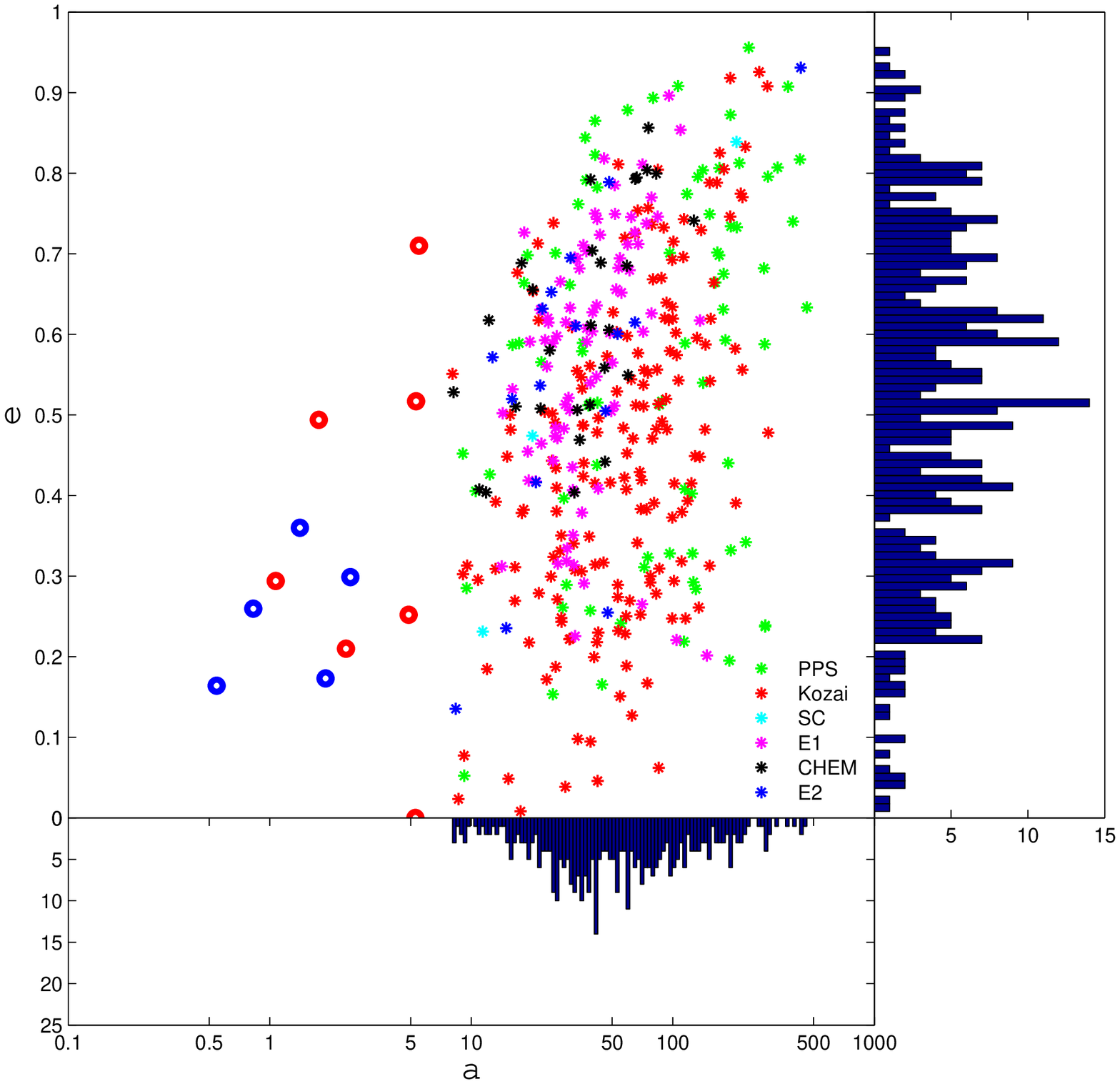}\\
  \caption{Scattering plot of semi-major axes and eccentricities of companions of HJ candidates in our simulations with ${a_{\rm in}}= 5$ au is shown in main diagram. We distinguish PPS, Kozai, CHEM, SC, E1, E2 using different colors.  Respective histograms of $a$, $e$ locate in bottom and right. The hollow blue circles represent the observed companions of HJs in ups And, Wasp-47 and HIP 14810, while the hollow red circles represent the others in Table \ref{tab:HJC}. }\label{fig:ae4}
\end{figure}

Among six high-eccentricity mechanisms to form HJs, only PPS can produce lonely HJs.  The occurrence of other mechanisms needs at least one companion of HJs. About $2.4\%$ of HJ candidates are alone in the group of ${a_{\rm in}}= 1$ au. That means most HJs should have companions. During orbital circularization of inner planets, part of orbital energy is dissipated and orbital angular momentum keeps constant roughly \citep{Wu2011}. The orbital elements of outer planets don't change a lot neglecting gas damping on $e$ and $i$.  We give scattering plot as well as histograms of semi-major axes and eccentricities of companions of HJ candidates in Fig. \ref{fig:ae2} and Fig. \ref{fig:ae4}. Different colors represent different high-eccentricity mechanisms to form HJ candidates.  The hollow blue and red circles represent the observed companions of HJs.

The semi-major axes and eccentricities of companions of HJ candidates formed through PPS and Kozai are most widely distributed. When ${a_{\rm in}}= 1$ au,  the semi-major axes of companions of HJ candidatesformed through PPS and Kozai distribute further than 1 au ( 8 au for ${a_{\rm in}}= 5$ au ).  The peak of semi-major axes of companions is around 5 au (45 au for ${a_{\rm in}}= 5$ au). While in the other five mechanisms,  the semi-major axes of companions distribute with relatively small ranges .  The occurrences of CHEM have strict requirements on the eccentricities of outer planets. The eccentricities of companions of HJ candidates formed through SC, E1 and E2 are usually moderate.

Among the observed HJs, there are 11 HJs having planetary companions. Among the companions, at least one is heavier than $0.3{M_J}$. The multi-planet systems including HJs are listed in Table \ref{tab:HJC}. Theses massive companions having both of semi-major axes and eccentricities detected are displayed in Fig. \ref{fig:ae2} and Fig. \ref{fig:ae4} by hollow circles. 



During disk migration, the damp of the eccentrities is one to two orders of magnitudes quiker than the semi-major axes \citep{Lee2002}. Suppose the companions of HJs could migrate to the current locations in gas-rich disks, the eccentricities shouldn't be so large as some observed companions, for example Pr0211 c and HD217107 c in Table \ref{tab:HJC}.  So, we speculate that a part of HJs observed in multi-planet systems form through high-eccentricity mechanisms. 


\section{CONCLUSIONS AND DISCUSSIONS}\label{sec:CandD}

\subsection{Conclusions}\label{sec:Con}

We made statistical analysis about the formation of HJ candidates through high-eccentricity mechanisms after dynamical instability of multi-planet systems with initial 2-5 Jupiter-like planets.

During classifying different high-eccentricity mechanisms to form HJ candidates, we found two new mechanisms which can produce high inclined even retrograde HJs.  In the first one, marked as E1, CHEM modulate the eccentricity of inner planet in a long period, while Kozai modulate a short period. Comparing with CHEM, this mechanism largely increase the inclination of inner planet largely, generally ${40^ \circ } \sim {80^ \circ }$, even retrograde orbits and raise the possibility of prograde high inclined HJs during tidal circularization comparing with Kozai-Lidov mechanism. In the second one, marked as E2, the orbit can flip between prograde and retrograde, which is well-coupled with outer orbit.

Then, we obtained the efficiencies of high-eccentricity mechanisms to produce HJ candidates.  Initial planet number, inner semi-major axes and space separation influence the possibility of HJ candidates formation.  In two-giant systems, $\le 1\%$ of cases can form prograde HJ candidates. If $N \ge 3$, the efficiencies to form HJ candidates are slightly sensitive to the planet number. The efficiencies are about $7\%  \sim 9\% $ for ${a_{\rm {in}}}= 1$ au, while $4\%  \sim 5\% $ for ${a_{\rm in}}= 5$ au.

In the system consisting of more than two planets, HJs forming through secular chaos are relatively easy \citep{Wu2011}.  However, in our simulations, the numbers of remained planets after dynamical instability are usually less than three. Secular chaos is not the main channel among high-eccentricity mechanisms for HJs formation within ${10^7}yr$. Kozai-Lidov mechanism playing a part in both Kozai and E1, is the most important high-eccentricity mechanism to produce HJs according to our simulations. 

\subsection{Discussions}\label{sec:Dis}
%
General relativity plays a complex role on the evolution of eccentricity of inner orbit. When the general relativistic timescales of precession of the arguments of pericenter are several orders of magnitude shorter than the Newtonian Kozai-Lidov timescales, the inner eccentricity will be suppressed. Otherwise, the eccentricity will be excited. Particularly, when the two timescales are comparable, the inner eccentricity will obtain resonant-like excitation around some critical values of the outer semi-major axis\citep{Naoz2013PN}.
In our simulations, the 1PN timescales (in relation to $a_{1}^{-2}$ as well as $a_{2}^{-2}$) are two orders of magnitude shorter than the timescales of Newtonian quadrupole terms in a ratio of $5\%$. In this situation, the inner eccentricity can be suppressed. Otherwise, the eccentricity will be excited mostly. So, if the general relativity was considered, the efficiencies of Kozai-Lidov mechanism is likely to increase. And the possibility of flipping orbit in CHEM mechanism may increase due to the excitation of inner eccentricity. The values and shortcomings of our results will be pointed out after the discussion of tidal effects.

Tidal dissipation can act on the semi-major axis as well as the angles between the spin axis of the star and orbital orientation of the planets. The changes of the semi-major axis depend on the stellar spin frequency and the planetary orbital frequency. Tidal dissipation can cause planetary orbits decay when the stellar spin frequency is lower than the planetary orbital frequency, and vice versa.  In CHEM mechanism, the tides may effectively halt the orbital flip, for the flipping needs extreme high eccentricity. The planets close to the host extremely may be disrupted due to tidal effects \citep{Li2014}. The disruption distance $\approx {f_t}{R_P}(m_2/(m_1+m_2))^
{-1/3}$,$ f_t=1.66 \sim 3.6$ \citep{Faber2005,Guillochon2011,Naoz2012,Petrovich2015b}.
Also the tidal effect can influence the final spin-orbital angles between HJs and the hosts, and shaping the obliquity distribution of exoplanets \citep{Valsecchi2014a,Valsecchi2014b}.

According to the observations, there is a temperature boundary $ \approx 6250$ K of the hosts which own highly inclined HJs \citep{Winn2010}. The inclinations of HJs with hot stars usually range from 0 to $\pi $. While in cold star systems, the orbital orientations of HJs are usually aligned with the spin axis of the hosts.  
There are two mechanisms may reduce the obliquity of HJs in cold star system. The tidal effects between the surface layer of cold star and HJs may decrease the inclinations to zero \citep{Albrecht2012,Lai2012}. The effect of magnetic breaking of solar type stars can also damp the inclinations of HJs \citep{Dawson2014}.  
Rogers \citep{Rogers2012,Rogers2013a,Rogers2013b} stated the spin axis of hot stars can be reorientated by internal gravity wave, give rise to the obliquity of HJs.  
So, simply setting a tidal boundary is not sufficient to be considered as an indicator for final circularization. We just considered the gravity because the aim of this article is to find out the relative efficiencies of different high-eccentricity mechanisms during planet-planet interactions. The efficiencies of HJs by high-eccentricity mechanisms will become clearer if the orbital circularization of planets including tidal effect and general relativity can be worked out. And the final distribution of obliquities of HJs will need more factors such as disk migration, the spin and physical structures of the planets as well as the magnetic field, the temperature and inner gravity waves of the host. 

From our simulations, we find the formation possibilities of HJs and retrograde ones is positively related to the number of giants. More giants need more massive planetary disk, as well as more massive and hotter central stars.  From this point, the formation efficiency of high inclined HJs  is further related to the temperature of star.

Up to now, there are 10 retrograde HJs observed in single star systems \footnote{http://www.astro.keele.ac.uk/jkt/tepcat/rossiter.html}, about $2.8\%$ of observed HJs.  There are 9 retrograde HJs with the temperature of star $\ge 6250$ K, about $13\%$ among the HJs in hot star systems.
In our simulations, when ${a_{\rm in}}= 1$ au, the lower limits of retrograde HJ candidates with $N \ge 3$ are about $2\%$, while about $8\%$ for ${a_{\rm in}}= 5$ au. 

If high-eccentricity mechanisms are the main routes to produce HJs, considering retrograde HJs during orbital flip as well as highly inclined HJs formed through such mechanisms of ``primordial misalignment" mentioned in Section \ref{sc:intro}, the propotion of observed retrograde HJs among the ones in hot star systems is reasonably greater than the lower limits in our simulations. 

%

Although due to detection limits of observation techniques, only 11 HJs are observed having planetary companions.
\citet{Knutson2014} searched for companions to these transiting planets with orbital periods between 0.7–11 days and masses between 0.06–11$M_{J}$ using the Keck HIRES instrument and found orbital semi-major axes of companions typically between 1–75 AU. According to comparisons between the retrograde proportion of HJs, as well as $a \sim e$ scatter diagram of planetary companions of HJs in our simulations and the observations, we can also find some clues about formation of HJs. We speculate there are at least two populations of HJs. One population migrate into the boundary of tidal effects due to the interaction with gas disk, such as these hot Jupiter systems, ups And, Wasp-47 and HIP 14810. They usually have more than two planets with lower eccentricities, and keep dynamical stable in relatively compact orbital configurations. Another population form through high-eccentricity migrations after disk migration. This kind of HJs usually has companions with moderate or high eccentricities in distant orbits. 
With developments of observation devices and methods, more HJs and their companions will be discovered, the characteristics of exo-planet systems with HJs will be clearer. Between disk migration and high-eccentricity migration, which one is the main channel to form HJs is still an open question.

\acknowledgements
Thanks Ji-Wei Xie for useful discussions and suggestions. Thanks donghong Wu for using software. Our research is supported by the Key Development Program of Basic Research of China (No. 2013CB834900), the National Natural Science Foundations of China (No. 11503009 , 11333002 and 1166116101),  985 Project of Ministration of Education and Superiority Discipline Construction Project of Jiangsu Province, ``Search for Terrestrial Exo-Planets,'' and Jiangsu Province Innovation for PhD candidate (No. CXLX13 032).

\newpage
\appendix

\section{list of multi-planet systems having hot Jupiters} \label{sec:tables}

\begin{table}[htbp]
\centering
\scalebox{0.7}{
\begin{tabular}{|c|c|c|c|c|c|c|c|}
\hline
Host Name&Letter&Method&Period [days]&a [au]&e &mass[$\rm M_J$]&Radius[$\rm R_J$]\\
\hline
ups And &b&RV&	$4.617033\pm0.000023$&	$0.05922166\pm0.00000020$&	$0.02150\pm0.00070$&	$0.6876\pm0.0044$&	\\
\hline
ups And &c&RV&	$241.258\pm0.064$&	$0.827774\pm0.000015$&$0.2596\pm0.0079$&	$1.981\pm0.019$	&\\
\hline
ups And &d&RV&	$1276.46\pm0.57$&	$2.51329\pm0.00075$&	$0.2987\pm0.0072$&	$4.132\pm0.029$&	\\
\hline
WASP-8 &b&Transit&$8.158715_{ - 0.000015}^{ + 0.000016}$&$0.0801_{ - 0.0016}^{ + 0.0014}$	&	$0.3100 _{ - 0.0024}^{ + 0.0029}$&	$2.244 _{ - 0.093}^{ + 0.079}$	&\\
\hline
WASP-8	&c&RV&		$4323 _{ - 380}^{ + 740}$	&$5.28_{ - 0.34}^{ + 0.63}$	&$0.0\pm0.0$	&$9.45_{ - 1.04}^{ + 2.26}$&	\\
\hline
WASP-47 &b&Transit&	$4.16071\pm0.00038	$&$0.052\pm0.011$	&$0.0028_{ - 0.0020}^{ + 0.0042}$	&$1.21_{ - 0.39}^{ + 0.59}$	&$1.15\pm0.25$\\
\hline
WASP-47 &c&RV&	$580.7\pm9.6$&	$1.41\pm0.30$	&$0.36\pm0.12$	&$1.57_{ - 0.59}^{ + 1.00}$	&\\
\hline
WASP-47 &d&Transit&	$9.095\pm0.014$	&$0.0880\pm0.0190$&	$0.00600_{ - 0.00410}^{ + 0.00980}$	&$0.0529_{ - 0.0220}^{ + 0.0378}$&	$0.331\pm0.073$\\
\hline
WASP-47	  &e&Transit&$	0.789636\pm0.000017	$&$0.0173\pm0.0038$&	$0.030_{ - 0.020}^{ + 0.036}$&	$0.0286_{ - 0.0113}^{ + 0.0173}$&	$0.162\pm0.036$\\
\hline
WASP-41 &b&Transit&	$3.0524040\pm0.0000009$&	$0.040\pm0.001$&$	<0.026$&	$0.94\pm0.05$	&$1.18\pm0.03$\\
\hline
WASP-41	  &c&RV&$	421\pm2$	&$1.07\pm0.03$	&$0.294\pm0.024$	&$3.18\pm0.20$	&\\
\hline
Pr0211    &b&RV&	$2.14610\pm0.00003$&	$0.03176\pm0.00015$&$0.011_{ - 0.008}^{ + 0.012}$		&$1.88\pm0.03$	&\\
\hline
Pr0211      &c&RV&	$4850_{ - 1750}^{ + 4560}$&	$5.5_{ - 1.4}^{ + 3.0}$	&$0.71\pm0.11$&$	7.79\pm0.33$&	\\
\hline
Kepler-424&b&Transit&	$3.3118644\pm0.00000039$& $0.044_{ - 0.004}^{ + 0.005}$ &		&$1.03\pm0.13$&	$0.89_{ - 0.06}^{ + 0.08}$\\
\hline
Kepler-424&	c&RV&	$223.3\pm2.1$&$0.73_{ - 0.07}^{ +0.08}$	&		&$6.97\pm0.62$&	\\
\hline
KELT-6  &b&Transit&	$7.8455821\pm0.0000070$&	$0.080\pm0.001$	&$0.029_{ - 0.013}^{ +0.016}$&	$0.442\pm0.019$	&$1.18\pm0.11$\\
\hline
KELT-6    &c&RV&$1276_{ -67	 }^{ +81}$	&$2.39\pm0.11$&	$0.210_{ -0.036 }^{ +0.039}$&	$3.71\pm0.21$&		\\
\hline
HIP 14810&b&RV&	$6.673855\pm0.000019$	&$0.0692\pm0.0040$	&$0.14270\pm0.00094$	&$3.88\pm0.32$	&\\
\hline
HIP 14810&c&	RV&	$147.730\pm0.065$&	$0.545\pm0.031$	&$0.164\pm0.012$	&$1.28\pm0.10$	&\\
\hline
HIP 14810&d&RV&	$952\pm15$&	$1.89\pm0.11	$&$0.173\pm0.037$	&$0.570\pm0.052$&\\
\hline
HD 217107&b&RV&	$7.126816\pm0.000039$&	$0.0748\pm0.0043	$&$0.1267\pm0.0052$	&$1.39\pm0.11$	&\\
\hline
HD 217107&c&RV	&$4270\pm220$&	$5.32\pm0.38$	&$0.517\pm0.033$	&$2.60\pm0.15$	&\\
\hline
HD 187123&b&RV&	$3.0965828\pm0.0000078$&	$0.0426\pm0.0025$&	$0.0103\pm0.0059$	&$0.523\pm0.043$&	\\
\hline
HD 187123&c&RV&	$3810\pm420$&	$4.89\pm0.53	$&$0.252\pm0.033	$&$1.99\pm0.25	$&\\
\hline
HAT-P-44&b&Transit&	$4.301219\pm0.000019$&	$0.0507\pm0.0007$&$	0.044\pm0.052$	&$0.352\pm0.029$	&$1.242_{ -0.051 }^{ +0.106}$\\
\hline
HAT-P-44&c&RV&	$872.2\pm1.7$&	$1.752\pm0.025$	&$0.494\pm0.081$	&$4.0_{ -0.8 }^{ +1.4}$	&\\
\hline
\end{tabular}}
  \caption{ The data are selected in website \url{https://exoplanetarchive.ipac.caltech.edu/}.}\label{tab:HJC}
\end{table}

\end{document}